\documentclass[proceedings]{JHEP3}
\usepackage{epsfig}                   

\PrHEP{PrHEP hep2001}                   
\conference{International Europhysics Conference on HEP}                

\def\qb{\overline{Q}}
\def\qqb{\overline{Q}\hspace{1pt}^2}
\def\als{\alpha_s}
\def\mev{\,{\rm MeV}}
\def\gev{\,{\rm GeV}}

\newcommand{\lsim}{\raisebox{-4pt}{$\,\stackrel{\textstyle
                                                         <}{\sim}\,$}}
\newcommand{\gsim}{\raisebox{-4pt}{$\,\stackrel{\textstyle
                                                         >}{\sim}\,$}}

\title{$\gamma^* \gamma^{(*)} \to \pi$ transitions and the pion 
       distribution amplitude}

\author{Carsten Vogt
        \thanks{Present address: Nordita, Blegdamsvej 17, DK-2100 Copenhagen,
         Denmark.}
        \thanks{Based on work together with M. Diehl and P. Kroll.} \\ 
        Fachbereich Physik, Universit\"at Wuppertal, 42097 Wuppertal, 
        Germany\\        
        E-mail: \email{cvogt@nordita.dk}}             

\abstract{We discuss what can be learned about the shape of the pion 
          distribution amplitude from the form factor for 
          $\gamma^*\gamma^{(*)}\to\pi$ transitions.}

\begin{document}

\section{Introduction}

The determination of hadronic distribution amplitudes by theoretical 
as well as by experimental means has been and still is an active area of 
research. The form factor $F_{\pi\gamma}$ for transitions 
$\gamma^* \gamma \to \pi$, for instance, has been discussed by many 
authors in order to find constraints for the shape of the pion distribution 
amplitude~\cite{jak96,mus97,bro98}, to name but a few.
The pion distribution amplitude is an important ingredient in, e.g., 
the electromagnetic pion form factor, pion pair production in 
two-photon collisions and, moreover, in exclusive B meson decays into 
pseudo-scalars, where strong interaction effects are currently under active 
investigation.

In this talk we discuss the transition form factor $F_{\pi\gamma^*}$ 
for doubly virtual photons~\cite{dkv01}, 
$\gamma^* \gamma^* \to \pi$, and explore what can be learned from it 
beyond what is already known from the case of the real-photon limit 
$\gamma^* \gamma \to \pi$.

\section{The transition form factor $F_{\pi\gamma^*}$}

We consider the case of space-like photon virtualities, denoted by
$Q^2=-q^2, \, Q'^2=-q'^2$, which is accessible in $e^+ e^- \to e^+ e^- \pi$. 
It is convenient to define
$\qqb  = (Q^2 + Q'^2)/2 , \, \omega= (Q^2 - Q'^2)/(Q^2 + Q'^2)$.
When at least one of the two photon virtualities is large compared to 
a hadronic scale the $\gamma^* \gamma^* \to \pi$ form factor factorises 
into a partonic hard scattering amplitude and a soft hadronic matrix 
element, which is parametrised by the universal pion distribution 
amplitude $\Phi_\pi$~\cite{lep79}.
The leading twist, next-to-leading order (NLO) $\als$ expression for 
$F_{\pi\gamma^*}$ is given by~\cite{agu81,bra83}:
\begin{equation}
 F_{\pi\gamma^*}(\qb,\omega) = \frac{f_\pi}{3\sqrt{2}\, \qqb}\,
  \int_{-1}^{\;1} {d} \xi\, \frac{\Phi_\pi(\xi,\mu_F)}{1-\xi^2\omega^2}\, 
  \left[1 + \frac{\als(\mu_R)}{\pi}\,{\cal K}(\omega,\xi,\qb/\mu_F) 
  \right] \, ,
\label{fpgvirtual}
\end{equation} 
The function ${\cal K}(\omega,\xi,\qb/\mu_F)$ is the NLO 
hard scattering kernel in the $\overline{\rm MS}$ scheme.
$\Phi_\pi(\xi,\mu_F)$ is the distribution amplitude of the pion's 
valence Fock state with $\xi= 2 x -1$ and $x$ being the light-cone 
momentum fraction of the quark. $\mu_F$ is the factorisation scale
and $\mu_R$ denotes the renormalisation scale, both of which are to 
be taken of order $\qb$. $f_\pi \approx 131\mev$ is the pion decay 
constant. The pion distribution amplitude can be expanded in terms
of the Gegenbauer polynomials $C_n^{3/2}(\xi)$, the eigenfunctions 
of the leading-order (LO) evolution kernel~\cite{lep79}:
\begin{equation}
 \Phi_{\pi}(\xi,\mu_F)=\Phi_{\rm AS}(\xi) 
  \left[1+ \sum^\infty_{n=2,4,...}B_n (\mu_F) \,C_n^{3/2}(\xi)\right] \,.
\label{evoleq}
\end{equation}
To lowest order, the scale dependent Gegenbauer coefficients $B_n$
are known to evolve according to 
\begin{equation}
 B_n (\mu_F)= B_n (\mu_0)\,
  \left(\frac{\als(\mu_F)}{\als(\mu_0)}\right)^{\gamma_n /\beta_0}\,,
\label{evolve}
\end{equation}
where $\mu_0$ is a reference scale which we choose to be 1 GeV and 
$\beta_0=11-2 n_f /3$. The anomalous dimensions $\gamma_n$ are positive
numbers increasing with $n$ such that for large scales any distribution 
amplitude evolves into the asymptotic form, given by 
\begin{equation}
\Phi_{\rm AS}(\xi)= \frac{3}{2} (1-\xi^2) \,.
\end{equation}
It immediately follows that in the real-photon case (i.e., $\omega = \pm 1$)
the transition form factor approaches the limit 
\begin{equation}
F_{\pi\gamma}(\qb,\omega=\pm 1) \longrightarrow  \frac{f_{\pi}}
{\sqrt{2}\, \qqb} 
\label{asy}
\end{equation}
as $\ln\mu_F$ becomes large. Note that this is a parameter-free QCD 
prediction, given that $f_\pi$ is known.

Using the Gegenbauer expansion~(\ref{evoleq}) the form 
factor~(\ref{fpgvirtual}) can be written as 
\begin{equation}
 F_{\pi\gamma^*}(\qb,\omega) = \frac{f_{\pi}}{\sqrt{2}\: \qqb} 
 \left[ c_0(\omega,\mu_R) 
 + \sum_{n=2,4,\ldots} c_n(\omega,\mu_R,\qb/\mu_F)\, B_n(\mu_F)
 \right] \,,
\label{f-two}
\end{equation}
with analytically computable functions $c_n$, which characterise the 
sensitivity of $F_{\pi\gamma^*}$ to the Gegenbauer coefficients $B_n$. 
The lowest coefficients $c_n$ are shown in Fig.~\ref{fig:coeffs}.
Here and in the following we use the two-loop expression of $\als$ with 
$n_f=4$, $\Lambda^{(4)}_{\overline{\rm MS}}= 305 \mev$ and we set 
$\mu_F=\mu_R=\qb$. We see that for $n\ge 2$ the functions $c_n$ show a 
very fast decrease as $\omega \to 0$, i.e., the transition form factor 
is sensitive to the coefficients $B_n$ only in the real-photon limit 
$\omega \to 1$. At $\omega=1$ we find 
\begin{equation}
 c_n(\omega=1) = 1+\frac{\als(\mu_R)}{\pi}\,{\cal K}_n(\omega=1,\qb/\mu_F)\,,
\end{equation}
\EPSFIGURE[h]{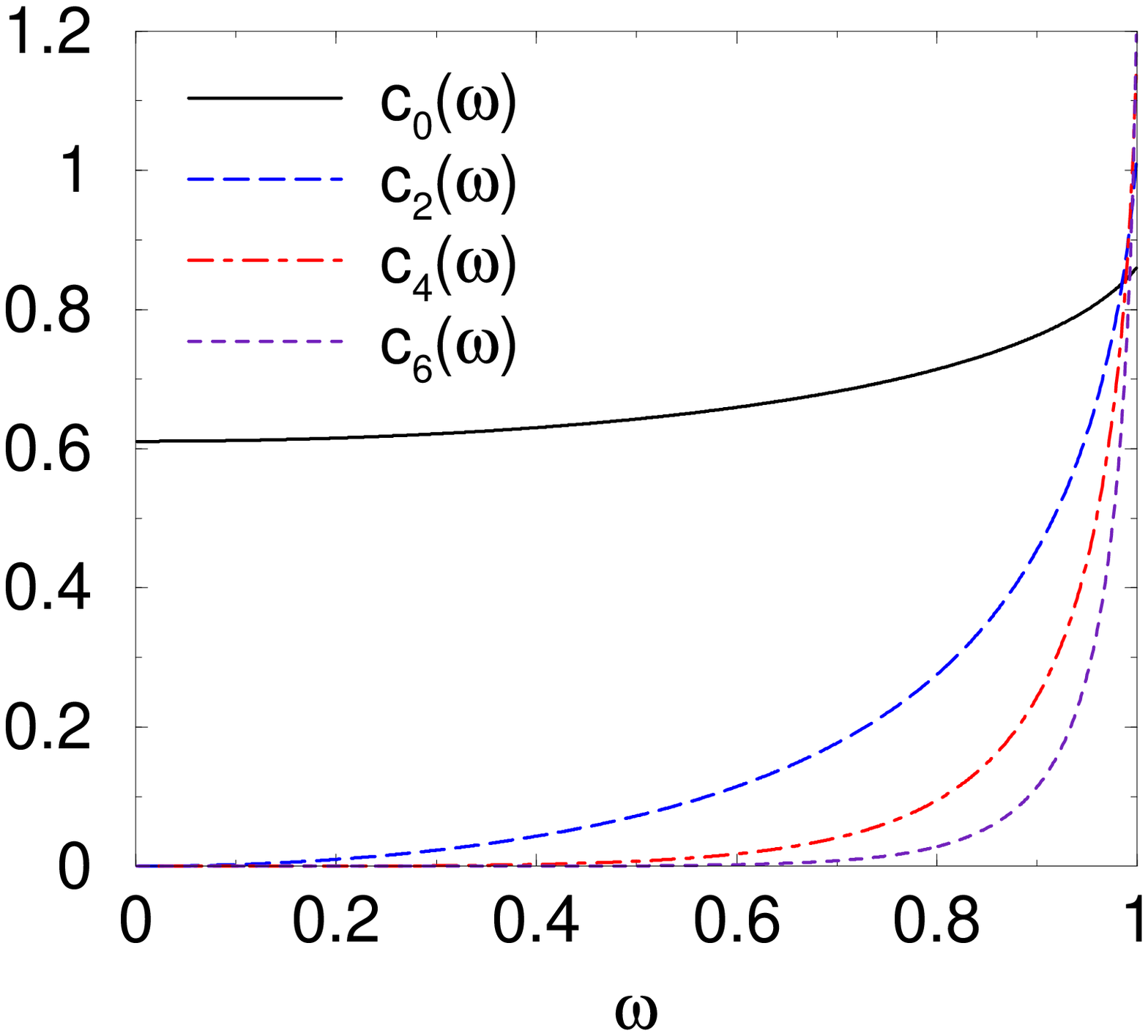,width=7cm}            
{The coefficients $c_n(\omega)$ in the expansion (\protect\ref{f-two}) 
of the $\gamma^*$--$\,\pi$ transition form factor. NLO corrections are 
included with $\mu_F=\mu_R=\qb$, which is taken as 2 GeV.\label{fig:coeffs}}
\noindent
which implies that the $\gamma$--$\pi$ transition form factor approximately 
probes the sum $1+\sum_n B_n$ of Gegenbauer coefficients. In order to 
extract information about individual Gegenbauer coefficients from the 
experimental data~\cite{cleo98} on the form factor one therefore
has to truncate the Gegenbauer series at some $n_0$ and assume $B_n=0$ 
for $n\ge n_0$. Taking $n_0=4$, for example, 
we find $B_2(\mu_0=1 \gev)=-0.06\pm 0.03$. Here we have restricted the 
experimentally available range of $Q^2$ to some $Q^2_{\rm min}$ between 
2 and 3 GeV$^2$, which is necessary to exclude large contributions from 
next-to-leading twist corrections. Further sources of uncertainty 
result from the experimental errors and from the fact that the choice of 
$\mu_F$ and $\mu_R$ is not unique at finite order of perturbation theory.

Allowing for nonzero $B_2$ and $B_4$ it is not possible to find unique
values for these coefficients since there is a linear correlation 
between $B_2$ and $B_4$, which is only slightly resolved due to a mild
logarithmic dependence and due to experimental errors.   
Performing a fit to the data with $Q^2_{\rm min}=2 \gev^2$ we find 
$B_2+B_4=-0.06\pm 0.08$ and $B_2-B_4=0.0\pm 0.9$ at $\mu_0=1\gev$. 
Thus, it might well be that the individual Gegenbauer coefficients 
are small and the pion distribution amplitude is close to its asymptotic 
form. On the other hand, with the presently available data on the 
$\gamma$--$\pi$ transition form factor, we cannot exclude large Gegenbauer
coefficients the sum of which being close to zero due to cancellations.

Apart from the above mentioned uncertainties, there are also power 
corrections which could spoil the phenomenological analysis and which 
we would like to briefly comment on. For $\omega$  close to 1, the
convolution~(\ref{fpgvirtual}) is sensitive to the endpoint regions
$\xi\to\pm 1$. This corresponds to the situation in which one of the
quarks in the pion has small momentum fraction such that soft effects
become important. In particular, corrections arising from partonic 
transverse momenta are non-negligible. In order to estimate these 
corrections, we follow the modified perturbative approach of 
Refs.~\cite{bot89,li92}. Here, the expression~(\ref{fpgvirtual}) is
replaced by a convolution of Fourier transforms of the pion 
light-cone wave function, the modified hard scattering kernel at LO $\als$ 
and the Sudakov factor, which accounts for resummed gluonic radiative 
corrections which are not contained in the wave function:
\begin{equation}
   F_{\pi\gamma^*}(\qb,\omega) = \frac{1}{4\sqrt{3}\pi^2} \int {d} \xi\,
     {d}^2{\bf b}\, \hat\Psi_{\pi}^*(\xi,-{\bf b},\mu_F)\,
     K_0(\sqrt{1 + \xi \omega}\: \qb\, b)\, 
     \exp\left[-S\left(\xi,b,\qb,\mu_R\right)\right]\,.
\label{fpgeq}
\end{equation}
Since the transverse separation $b$ acts as an infrared cut-off, 
the factorisation scale $\mu_F$ is set equal to $1/b$. As the
renormalisation scale $\mu_R$ we take the largest mass scale 
occuring in the propagator of the internal quark,
$\mu_R = \max{\{1/b, \sqrt{1+\xi\omega}\:\, \qb,
\sqrt{1-\xi\omega}\:\, \qb\}}$~\cite{li92}. Following~\cite{lep83,jak93} 
we assume for the light-cone wave function in $b$-space the simple form
\begin{equation}
\hat{\Psi}_\pi(\xi,{\bf b})=\frac{2\pi f_\pi}{\sqrt{6}}\, 
       \Phi_{\rm AS}(\xi) 
       \exp{\left[-\frac{ \pi^2 f_\pi^2}{2} (1-\xi^2)\, b^2\right]}
\label{modwf}
\end{equation}
in our estimate. The prediction of the $\gamma$--$\pi$ transition form 
factor in the modified perturbative approach using this wave function
leads to very good agreement with the CLEO data~\cite{kro96}.   

\DOUBLEFIGURE[t]{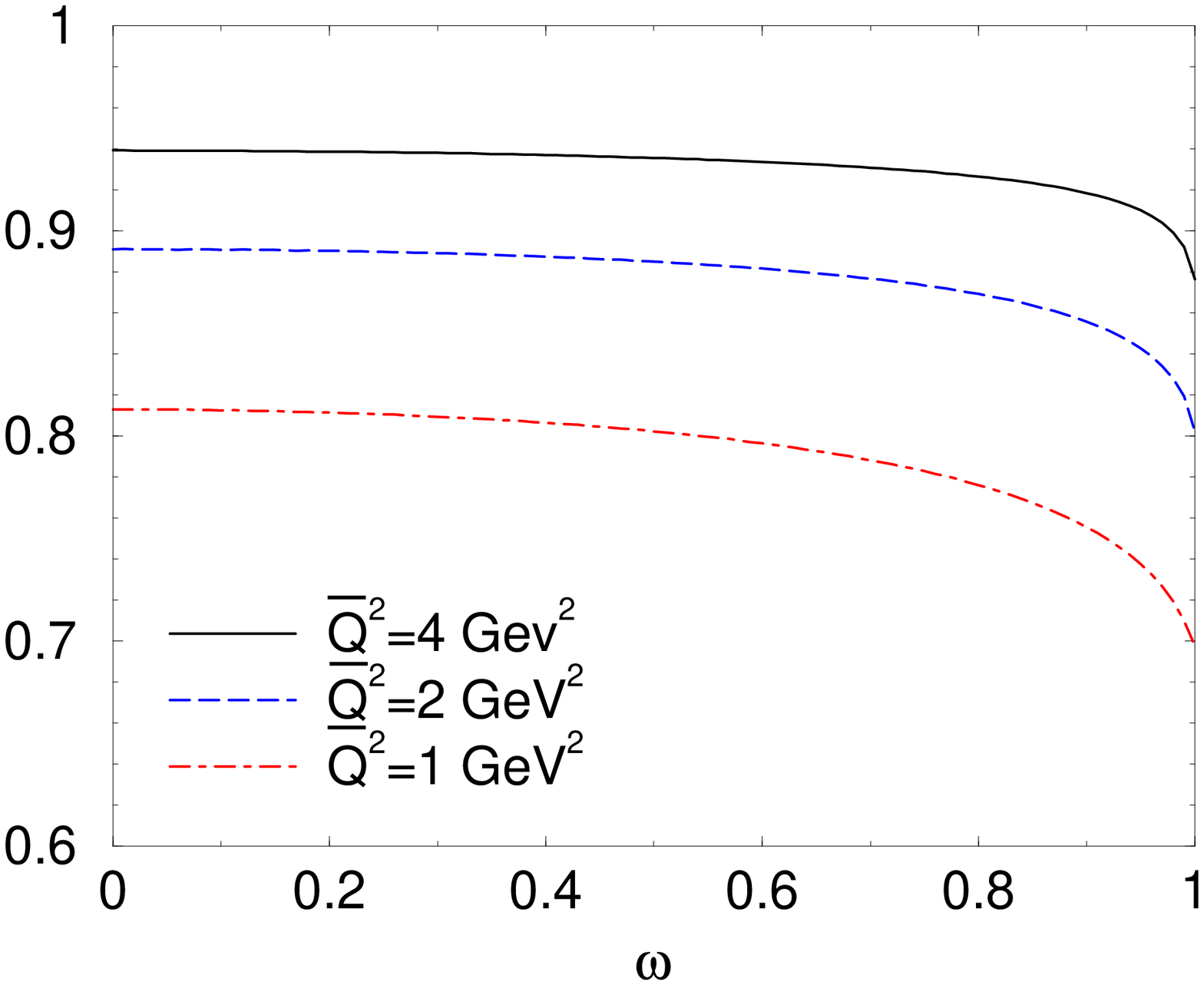,width=7cm}{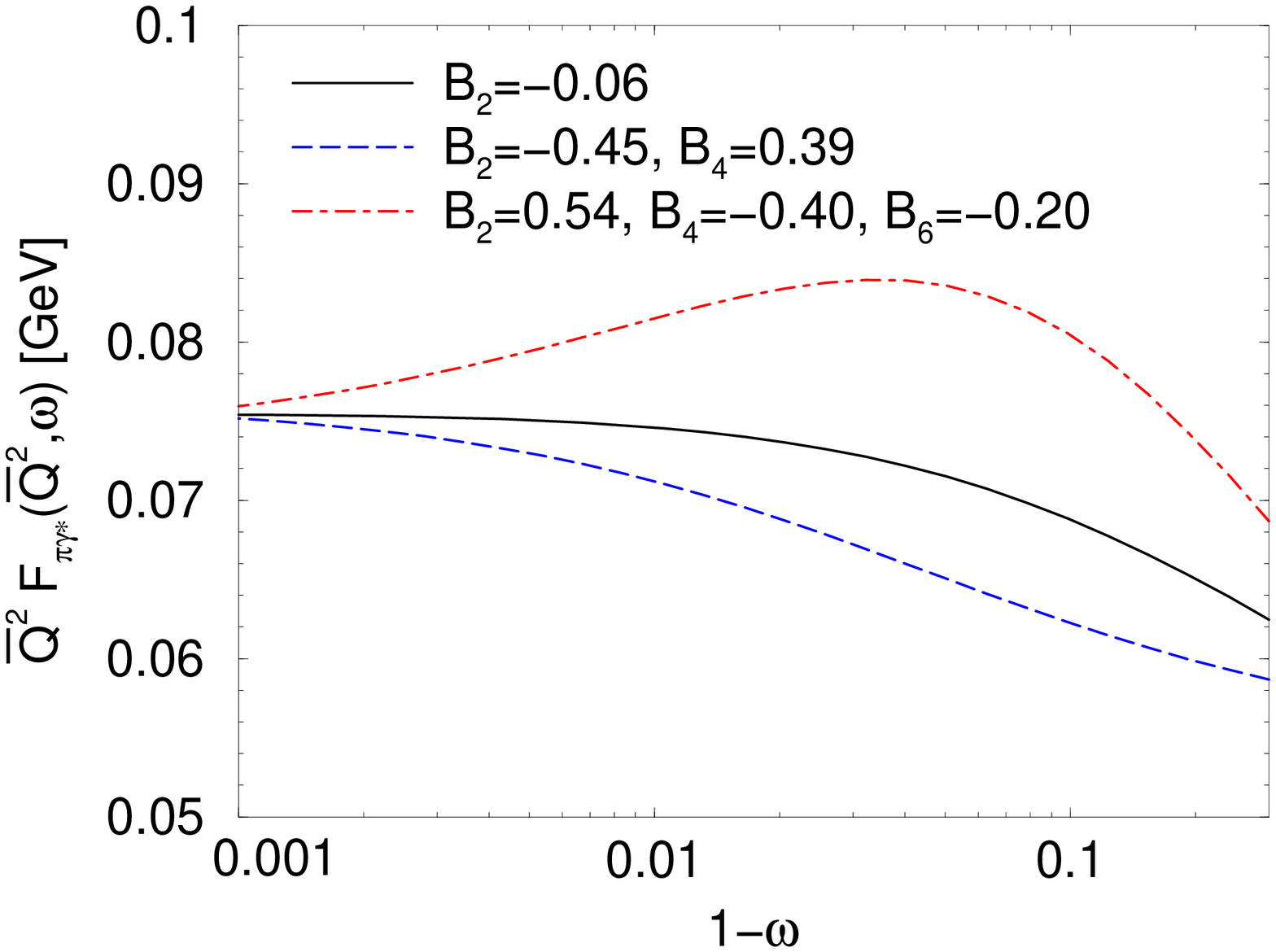,width=76mm} 
{Ratio of $F_{\pi\gamma^*}(\qb,\omega)$ in the modified
perturbative approach and in the LO leading-twist approximation at
$\qqb= 4\gev^2$ (solid line), $\qqb=2\gev^2$ (dashed line), and 
$\qqb= 1\gev^2$ (dash-dotted line). Here we have used the wave function
(\protect\ref{modwf}) in the modified perturbative approach and the asymptotic
pion distribution amplitude.
\label{fig:ratio}}
{The scaled form factor $\qqb F_{\pi\gamma^*}(\qb,\omega)$ calculated 
to NLO in the leading-twist approximation at $\qb^2=4 \gev^2$, using 
sample distribution amplitudes. The values of the Gegenbauer
coefficients are quoted at the scale $\mu_0 = 1 \gev$.  
All higher order Gegenbauer coefficients are taken to be zero.
\label{fig:endplot}}   
In Fig.~\ref{fig:ratio} we show the ratio of the predictions of the 
form factor in the modified perturbative approach and in 
leading-twist approximation at LO in $\als$. In both approximations
we employ the asymptotic form of the pion distribution amplitude. 
We see that the power
corrections are less than about 10\% for $\qb^2=4 \gev^2$.
Note that the Sudakov corrections amount to no more than 1.5\%
such that it is sufficient to retain only the leading logarithmic 
terms in the Sudakov function $S$ as given in Ref.~\cite{bot89}.

We have also checked that the results of Fig.~\ref{fig:ratio} 
essentially remain unchanged for $\omega\lsim 0.8$ when we include 
an effective mass $m_{\rm eff}=0.33\gev$ in the light-cone wave 
function~(\ref{modwf}) with appropriately adjusted parameters,
as proposed in Ref.~\cite{lep83}, for example. 
For $\omega\gsim 0.8$ the ratio then becomes up to 10\% smaller, 
since the inclusion of the mass term leads to a stronger suppression 
of the endpoint regions in the modified perturbative approach.

Although Fig.~\ref{fig:coeffs} clearly shows that in practice 
one cannot gain new informations on the Gegenbauer coefficients from 
$\gamma^*\gamma^*\to\pi$ transitions in a wide range of $\omega$,
one can nevertheless use the region where $\omega$ is close to but
different from 1 to obtain valuable informations beyond what is already known
from the real-photon limit. This is demonstrated in Fig.~\ref{fig:endplot}, 
where we plot the form factor for different choices of distribution 
amplitudes.
The kinematical range shown allows for a discrimination between distribution 
amplitudes with small individual Gegenbauer coefficients and 
distribution amplitudes with large $B_n$ the sum of which being small
in compliance with the constraint from the real-photon limit.    

We now turn to the kinematical region where $\omega$ significantly differs 
from 1. The fast decrease of the functions $c_n$ can be understood by
expanding the hard scattering kernel in Eq.~(\ref{fpgvirtual}) for small
$\omega$. Remarkably, one finds that each coefficient $B_n$ is suppressed
by a corresponding factor $\omega^n$. Neglecting terms of 
${\cal O}(\omega^4)$ we obtain
\begin{eqnarray}
 F_{\pi\gamma^*}(\qb,\omega)&=&\frac{\sqrt{2}f_\pi}{3\: \qqb} \left[
    1 - \frac{\als}{\pi} + 
  \frac15\, \omega^2 \left( 1-\frac53\frac{\als}{\pi} \right)
  \phantom{\frac{\qqb}{\mu_F^2}} \right. 
\nonumber \\
&&  \hspace{2em} \left. {}+ \frac{12}{35}\, \omega^2 B_2(\mu_F) 
    \left( 1 + \frac{5}{12} \frac{\als}{\pi} 
    \left\{ 1-\frac{10}{3} \ln\frac{\qqb}{\mu_F^2} \right\}
    \right) \right]
+ {\cal O}(\omega^4,\als^2) . \;\;
\label{fpgapprox}
\end{eqnarray}
For $\omega \to 0$ we thus have a parameter-free prediction from QCD to 
leading-twist accuracy, which is even valid over a wide range of $\omega$:
\begin{equation}
F_{\pi\gamma^*}(\qb,\omega)=\frac{\sqrt{2}f_\pi}{3\: \qqb}
         \bigg[ 1-\frac{\als}{\pi} \bigg]+ {\cal O}(\omega^2,\als^2) \,.
\label{qcd-pre}
\end{equation}
To LO $\als$, this result is known since long, 
see Ref.~\cite{cor66}. The $\als$-correction to the leading term
can be found in Ref.~\cite{agu81}. Any observed deviation from the 
leading-twist prediction would be a signal for large power corrections
and therefore, this prediction well deserves experimental verification.
For small $\omega$, the relation (\ref{qcd-pre}) has a status
comparable to the famous expression of the cross section ratio $R =
\sigma(e^+ e^-\to {\rm hadrons}) / \sigma(e^+e^-\to \mu^+ \mu^-)$.

%
%



\begin{thebibliography}{99}

\bibitem{jak96} R.\ Jakob, P.\ Kroll and M.\ Raulfs,
	        \jphg{22}{1996}{45}.
\bibitem{mus97} I.V.\ Musatov and A.V.\ Radyushkin, 
	        \prd{56}{1997}{2713}.
\bibitem{bro98} S.J.\ Brodsky, C.-R.\ Ji, A.\ Pang and D.G.\ Robertson,
                \prd{57}{1998}{245}.
\bibitem{dkv01} M.\ Diehl, P.\ Kroll and C.\ Vogt, Eur. Phys. J. C, in press
                [hep-ph/0108220].
\bibitem{lep79} G.P.\ Lepage and S.J.\ Brodsky,
                \prd{22}{1980}{2157}.                
\bibitem{agu81} F.\ Del Aguila and M.K.\ Chase,
                \npb{193}{1981}{517}.
\bibitem{bra83} E.\ Braaten, 
	        \prd{28}{1983}{524}.
\bibitem{cleo98} J.\ Gronberg {\em et al.}, CLEO collaboration,
	         \prd{57}{1998}{33}.
\bibitem{bot89} J.\ Botts and G.\ Sterman,
	        \npb{325}{1989}{62}.
\bibitem{li92}  H.-N.\ Li and G.\ Sterman,
	        \npb{381}{1992}{129}.
\bibitem{lep83} S.~J.~Brodsky, T.~Huang and G.~P.~Lepage, 
                {\it Particles and Fields}, Vol.~2, edited by A.~Z.~Capri 
                and A.~N.~Kamal, Banff Summer Institute, 1981,
                (Plenum, New York, 1983), p.~143.
\bibitem{jak93} R.~Jakob and P.~Kroll,
	        \plb{315}{1993}{463}.
\bibitem{kro96} P.\ Kroll and M.\ Raulfs,
                \plb{387}{1996}{848}.
\bibitem{cor66} J.M.\ Cornwall,
	        \prd{16}{1966}{1174}; \\
                G.\ K\"opp, T.F.\ Walsh and P.\ Zerwas, 
                \npb{70}{1974}{461}.

\end{thebibliography}
\end{document}